# Lamb waves sensor in liquid media utilizing higher-order quasi-longitudinal $S_5$ and $S_6$ modes

Muhammad Hamidullah, Nassim Rezzag, Céline Élie-Caille and Thérèse Leblois
FEMTO-ST Institute, CNRS UMR-6174,
University Bourgogne Franche-Comté, 25000 Besançon, France

*Abstract*— Higher-order quasi-longitudinal (QL) Lamb waves have been reported for sensor application in liquid media at a higher thickness-to-wavelength (h/λ) ratio than the fundamental symmetric modes, with the main advantage of having higher operating frequency and reduction in the total size of the device without reducing the plate thickness. However, the trade-off in the reduction of the electromechanical coupling coefficient ($K^2$) is a hinder in utilizing QL Lamb waves with higher-order numbers. Here we performed finite element simulation and verified experimentally the possibility of utilizing $S_5$ and $S_6$ QL Lamb wave modes in liquid media based on a two-port delay line device on a thick GaAs substrate. While the $K^2$ values are lower for QL modes at higher-order numbers, the signal amplitude of $S_{21}$ transmission parameter at the resonant frequency is still distinguishable from the noise floor. The result demonstrates the feasibility of higher-order modes at higher h/λ ratio for sensor application in liquid media, thus allowing further reduction in the wavelength and the total size of the sensor device.

## I. Introduction

Plate acoustic waves (PAW) with two free surfaces have an advantage over surface acoustic waves (SAW) with only one free surface for sensor application in liquid media [1]. In PAW, the sensing surface can be separated from the opposite surface with interdigital transducer (IDT), thus PAW-based sensors do not require an extra protection layer for the IDT to prevent contact with liquid. Several types of in-plane polarised PAW for sensors in liquid media have been reported, such as shear-horizontal acoustic plate modes (SH-APM) [2] and Lamb waves (LW) [3].

Fundamental antisymmetric ($A_0$) and symmetric ($S_0$) modes of LW are commonly used for sensors in liquid media without significant radiation of acoustic energy into the liquid under specific conditions. For the $A_0$ modes, the phase velocity ($v_{ph}$) must be lower than the sound velocity of the liquid so that the acoustic energy is not converted into longitudinal pressure waves in the liquid. In the case of $S_0$ modes, the particle displacement components must be a dominantly longitudinal component ($u_1$) with close to zero shear vertical ($u_3$) component. Both conditions are achieved at a low plate thickness-to-wavelength ((h/λ) ratio, typically below 0.1). Therefore, for $A_0$ and $S_0$ modes sensor in liquid media, the reduction in the plate thickness is necessary which causing the device to be fragile, or by using a larger wavelength with a trade-off in operating frequency and sensitivity reduction.

The limitations of low h/λ ratio for $A_0$ and $S_0$ modes were solved by using higher-order symmetric modes with quasi-longitudinal polarisation (QL-$S_n$). Anisimkin et al [4] reported QL-Sn modes in ST-cut quartz for a sensor in liquid media utilizing higher-order symmetry mode at h/λ ratio of 1.485. Verona et al [5] reported the occurrence of QL-$S_n$ modes in layered thin film ZnO/Si-substrate/ZnO up the 4th higher-order symmetric modes ($S_4$) with h/λ ratio of 2.6 [5]. Further reduction in the wavelength and the device size is an advantage of using symmetric modes with a higher order number, however, the electromechanical coupling coefficient ($K^2$) reduces as the order number increases. For two-port delay line device implementation, the reduction of $K^2$ will affect the signal strength received in receiver IDT and reduces the signal to noise ratio. In this paper, we perform finite element method (FEM) simulation to compare the performances of the first 6 higher-order symmetric QL-LW on GaAs substrate and we verify experimentally the feasibility of using $S_5$ and $S_6$ modes for sensor application in liquid media.

## II. FINITE ELEMENT SIMULATION

The occurrence of higher-order QL-LW modes at (100)-cut GaAs plate propagating along the <110> direction is previously reported for $S_1$ and $S_2$ modes at h/λ ratio of 0.6 and 1.2, respectively [6], where the $v_{ph}$ of higher-order symmetric modes ($S_n$) are close to the $v_{ph}$ of longitudinal bulk acoustic wave (LBAW). Due to the periodicity on the dispersion curves for higher-order modes, the h/λ ratio of quasi-longitudinal symmetric modes at order number n (QL-$S_n$) can be approximately obtained by multiplying the order number n with the factor of 0.6. Thus h/λ ratio for the first 6 higher-order QL modes are 0.6, 1.2. 1.8, 2.4, 3 and 3.6.

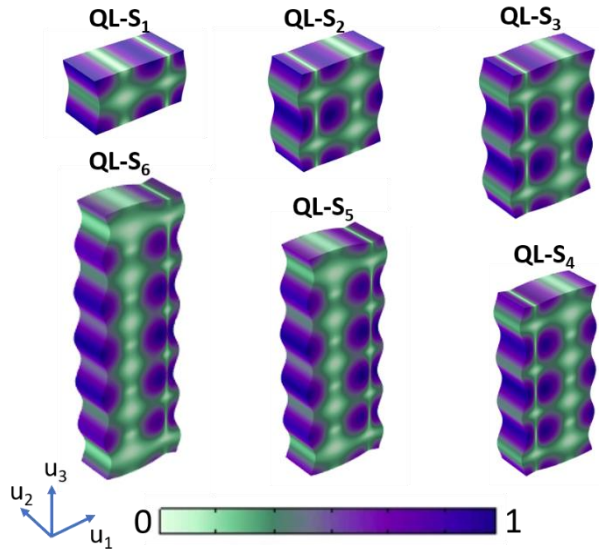

Fig. 1. Normalised total displacement profile of the first six higher-order quasi-longitudinal symmetric modes (QL-$S_n$) on (100)-cut GaAs propagating on <110> direction.

Figure 1 shows the total displacement profile of the first 6 QL-$S_n$ modes on GaAs plate obtained by COMSOL Multiphysics eigenfrequency simulation with the wavelength of 168 µm and the plate thickness of 100 µm, 200 µm, 300 µm, 400 µm, 500 µm and 600 µm for QL-$S_1$, QL-$S_2$, QL-$S_3$, QL-$S_4$, QL-$S_5$ and QL-$S_6$ modes respectively. As shown in figure 1, the total particle displacement profiles at the surface for all six modes are dominantly longitudinal.

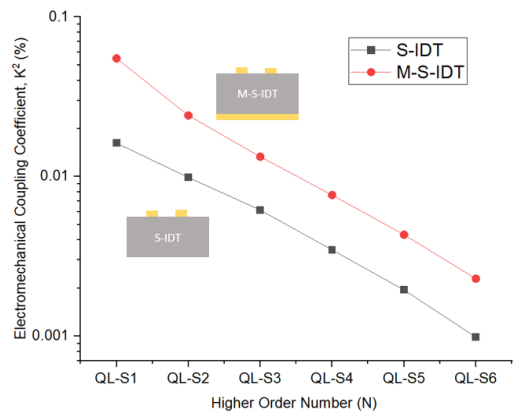

Fig. 2. $K^2$ of the first six QL-$S_n$ modes with S-IDT and M-S-IDT coupling configuration

In PAW with two free surfaces, the electrical boundary condition (BC) of the surfaces affects the value of $K^2$. The electrically shorted BC is achieved by the deposition of a metal layer on the opposite side of the IDT. Metal-substrate-IDT (M-S-IDT) and substrate-IDT (S-IDT) are electrical coupling configurations with and without additional metal layers respectively. The $K^2$ of the first six QL-modes are shown in figure 2 for two different coupling configurations. As shown in figure 2, the $K^2$ decreases as the order number increases. The $K^2$ of $S_6$ mode, for instance, is one order lower than the $K^2$ of $S_1$ mode which will be equal to 10 dB reduction in the signal strength. Furthermore, the M-S-IDT configuration slightly improves the $K^2$ to around two times higher than the $K^2$ value of the S-IDT configuration.

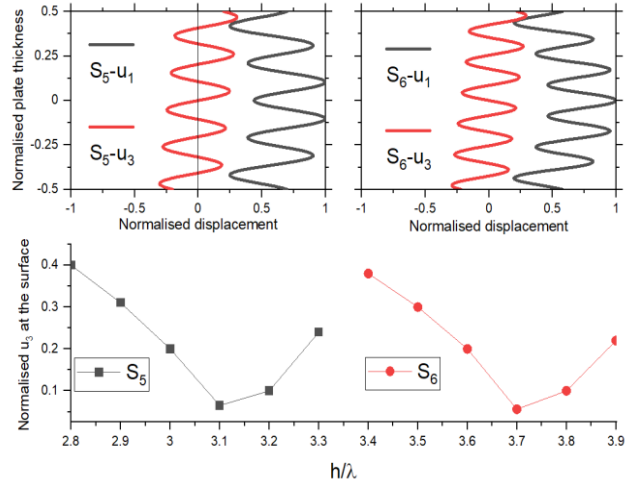

Fig. 3. (a) Particle displacement component profile of $S_5$ and $S_6$ modes and (b) Normalised shear vertical ($u_3$) displacement at the surface of the plate

Figure 3a shows the particle displacement components: longitudinal ($u_1$) and shear vertical ($u_3$) for $S_5$ and $S_6$ modes at $h/\lambda$ of 3 and 3.6 respectively. The shear-horizontal ($u_2$) displacement component is zero across the plate. As shown in figure 3a, the longitudinal displacement component is dominant across the plate for both modes. At the surface of the plate, the $u_3$ is around 20% of the maximum $u_1$ inside the plate. Figure 3b shows the normalised $u_3$ at the surface at $h/\lambda$ ratio of 2.8 to 3.2 for S5 mode and 3.3 to 3.9 for S6 mode. As shown in figure 3b, the lowest $u_3$ component is at $h/\lambda$ of 3.1 for $S_5$ and 3.7 at $S_6$ with normalised $u_3$ of 6.5% and 5.6% respectively. At these ratios, the lowest insertion loss is expected when the surface of the plate is in contact with liquid.

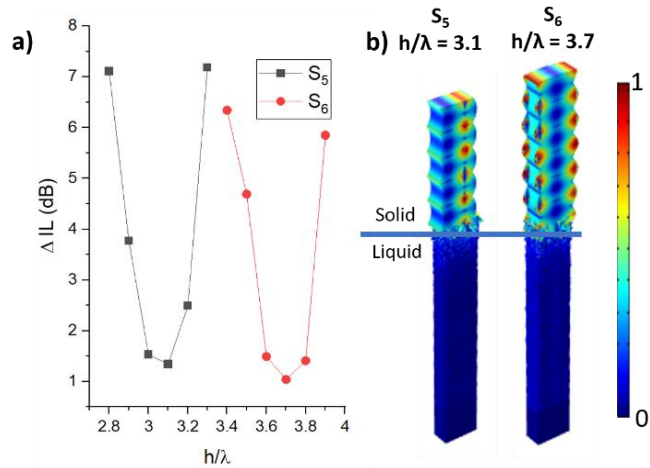

Fig. 4. (a) The insertion loss due to liquid damping for delay line with a centre to centre distance of 250λ and at $h/\lambda$ ratio of 2.8 to 3.2 for S5 mode and 3.3 to 3.9 for S6 mode and (b) Displacement profile of $S_5$ and $S_6$ modes with $h/\lambda$ ratio of 3.1 and 3.7 respectively.

Figure 4a shows the insertion loss simulation result when one surface of the plate is in contact with DI water for $h/\lambda$ ratio of 2.8 to 3.2 for S5 mode and 3.3 to 3.9 for S6 mode. Figure 4b shows the 3D total displacement profile of the $S_5$ and $S_6$ modes when the plate is in contact with the liquid. As expected, the lowest insertion loss is achieved at $h/\lambda$ of 3.1 for $S_5$ mode and at 3.7 for $S_6$ mode. For delay line with the centre to centre distance between transmitter and receiver IDT of 250λ, the minimum theoretical insertion loss due to the liquid damping is 1.34 dB for S5 mode and 1.04 dB for S6 mode. Furthermore, there is a margin of +/- 0.1 in h/λ ratio with relatively low insertion loss.

III. EXPERIMENTAL RESULTS

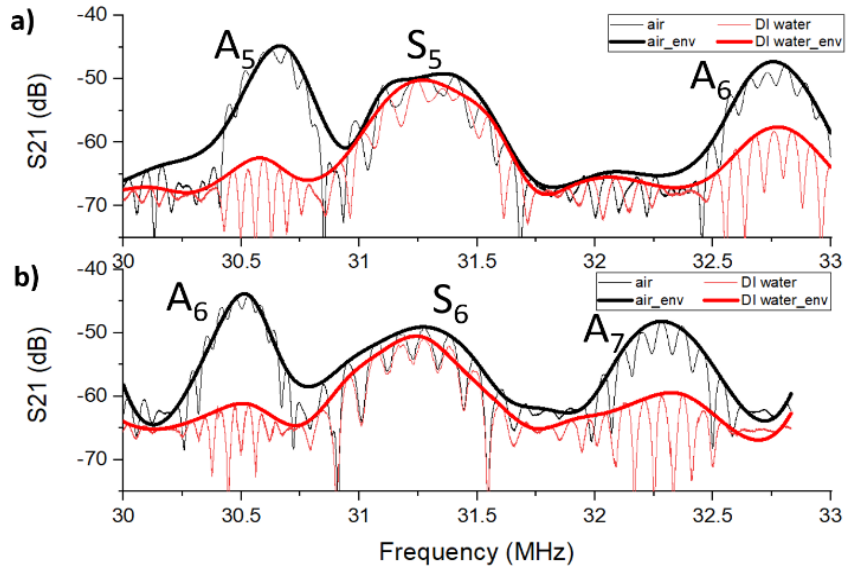

Fig. 5. Experimental result of two-port delay line device utilizing (a) $S_5$ mode and (b) $S_6$ mode in dry and liquid media

Two-port delay line devices were fabricated on (100)-cut GaAs wafer with the thickness of 500 μm and 625 μm and thickness variation of +/- 25 μm. For both wafers, the design of the IDTs is identical and the wavelength of the IDTs are 168 μm. Considering the thickness variation of the wafer, the h/λ ratios are ranging from 2.82 to 3.12 for $S_5$ mode and 3.57 to 3.86 for $S_6$ mode. The waves propagate at <110> direction parallel to the wafer flat. The IDTs have 50 pairs of source and ground with an aperture of 50λ. The delay length is 200λ, thus the centre-to-centre distance from the IDT receiver to the IDT transmitter is 250λ or equal to 42 mm. The metal electrode of 100 nm Al with 10nm Cr adhesion layer was deposited by a lift-off process. Gold metal layer was deposited at the back of the substrate to obtain the M-S-IDT coupling configuration. Furthermore, the metal layer at the sensing surface will neutralise the effect of liquid permittivity on the sensor responses. The GaAs wafers are attached to PCBs with a circular hole so that the delay line structures and the IDTs are suspended. 4194A Impedance/Gain-Phase Analyzer was used to obtain the frequency response and the power received in the IDT receiver normalised to the power sent from the IDT transmitter ($S_{21}$ scattering parameter).

The $S_{21}$ frequency responses of the sensor devices with the thickness of 500 μm and 625 μm are shown in figure 5a and 5b respectively. As shown in figure 5a, there are three resonant frequency peaks for the frequency range from 30 MHz to 33 MHz which correspond to $A_5$, $S_5$ and $S_6$ modes when the device is in a dry environment. However, when the surface of the plate is in contact with DI water, the reduction in $S_{21}$ signals are significant for $A_5$ and $A_6$ modes while the reduction in the $S_5$ mode is only around 1 dB. Similarly, in figure 5b, there are three resonant frequency peaks in dry environment but only the $S_6$ mode peak is observable when the plate is in contact with DI water with around 1.4 dB reduction in the $S_{21}$ signal. The $S_{21}$ signals of around -50 dB in contact with DI water for both $S_5$ and $S_6$ modes are distinguishable from the noise floor of -65 dB, thus showing the feasibility of those modes for sensor application in liquid media.

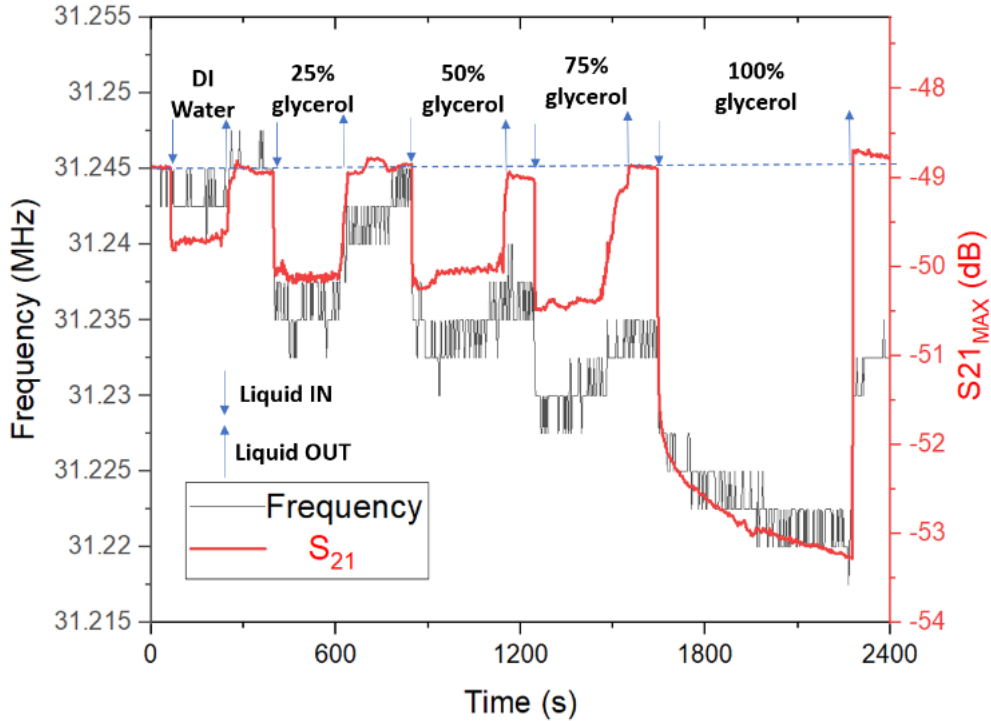

Fig. 6. Experimental results of QL-S5 mode as viscosity sensor with different concentrations of DI water/glycerol mixture

LabVIEW data acquisition software was used to evaluate the fluctuation in frequency and $S_{21}$ amplitude at the resonant frequency of the $S_5$ mode device. DI water/glycerol mixtures with different glycerol concentrations were used to evaluate the change in the signals under different liquid viscosity and the result is shown in figure 6. As shown in figure 6, both $S_{21}$ amplitudes and resonant frequencies reduce as the glycerol concentration increases. The results agree with the previously reported result of PAW viscosity sensors such as SH-APM [7]. The $S_{21}$ amplitudes return to the baseline amplitude after the device is no longer in contact with liquid, however, the frequency is remaining in the lower value, especially after the DI water/glycerol mixture was in contact with the plate surface. One possible explanation is the trace of glycerol remains in the plate surface act as a mass loading which reduces the resonant frequency. Thus, the $S_{21}$ amplitude is a more reliable signal than the frequency for viscosity sensor application.

## IV. CONCLUSION

The feasibility of utilizing quasi-longitudinal modes at higher-order numbers (QL-$S_5$ and QL-$S_6$) for a sensor in liquid media is reported in this paper using (100)-cut GaAs substrate. The employment of modes with higher-order numbers allows the reduction of wavelength thus increases the resonant frequency without reducing the thickness of the plate and improves the robustness of the device. While the reduction in $K^2$ is inevitable, the $S_{21}$ amplitude at the resonant frequency is distinguishable from the noise floor. The improvement in $K^2$ can be achieved by depositing highly piezoelectric thin film such as ZnO on GaAs [6] or by using plate substrate material with higher piezoelectricity constant.

## ACKNOWLEDGEMENT

This conference proceeding is part of SmOoC project that has received funding from the European Union's Horizon 2020 research and innovation programme under the Marie Skłodowska-Curie grant agreement No. 844135. The work was partly supported by the French Renatech network and its FEMTO-ST technological facility